\documentclass[]{spie}  
\usepackage[]{graphicx}
\usepackage{amsmath, amssymb}

\newcommand\farcs{\mbox{$.\!\!^{\prime\prime}$}}

\title{Scientific Design of a High Contrast Integral Field Spectrograph
for the Subaru Telescope}
\author{Michael W. McElwain\supit{a}, Timothy D. Brandt\supit{b}, Markus 
Janson\supit{b}, Gillian R. Knapp\supit{b}, Mary Anne Peters\supit{b},
Adam Burrows\supit{b}, Alexis Carlotti\supit{b}, Michael A. Carr\supit{b},
Tyler Groff\supit{b}, James E. Gunn\supit{b},
Olivier Guyon\supit{c}, Masahiko Hayashi\supit{d},
N. Jeremy Kasdin\supit{b}, Masayuki Kuzuhara\supit{e}, 
Robert H. Lupton\supit{b}, Frantz 
Martinache\supit{c}, David Spiegel\supit{f}, Naruhisa Takato\supit{c},
Motohide Tamura\supit{d}, Edwin L. Turner\supit{b}, Robert J.
Vanderbei\supit{b}
\skiplinehalf
\supit{a}Goddard Space Flight Center, Greenbelt, MD, USA; \\
\supit{b}Princeton University, Princeton, NJ, USA; \\
\supit{c}Subaru Headquarters, National Astronomical Observatory of Japan,
Hilo, HI, USA;\\
\supit{d}National Astronomical Observatory of Japan, Tokyo, Japan;\\
\supit{e}University of Tokyo, Tokyo, Japan;\\
\supit{f}Institute for Advanced Study, Princeton, NJ, USA.\\
}

\authorinfo{Further author information: Send correspondence to 
Michael McElwain\\email: mcelwain@princeton.edu}
 
\begin{document} 
\maketitle 

\begin{abstract}
Ground-based telescopes equipped with adaptive-optics (AO) systems and
specialized science cameras are now capable of directly detecting
extrasolar planets. We present the expected scientific capabilities of
CHARIS, the Coronagraphic High Angular Resolution Imaging
Spectrograph, which is being built for the Subaru 8.2 m
telescope of the National Astronomical Observatory of Japan. CHARIS
will be implemented behind the new extreme adaptive optics system at
Subaru, SCExAO, and the existing 188-actuator system AO188. CHARIS
will offer three observing modes over near-infrared wavelengths from
0.9 to 2.4 $\mu$m (the $y$-, $J$-, $H$-, and $K$-bands), including a
low-spectral-resolution mode covering this entire wavelength range and
a high-resolution mode within a single band. With these capabilities,
CHARIS will offer exceptional sensitivity for discovering giant
exoplanets, and will enable detailed characterization of their
atmospheres. CHARIS, the only planned high-contrast integral field
spectrograph on an 8m-class telescope in the Northern Hemisphere, will
complement the similar instruments such as Project 1640 at Palomar, and 
GPI and SPHERE in Chile.
\end{abstract}


\section{INTRODUCTION}
\label{sec:intro} 
The past seventeen years have seen the discovery of over 700 planets
around stars other than the Sun, commonly called ``exoplanets'' (see
\verb|www.exoplanets.org|).  Most of these exoplanets have been
discovered by the radial velocity variations they induce in their host
stars, or through periodic stellar dimming due to planetary transits.
Both methods are most sensitive to planets on short-period orbits.  The
magnitude of a radial velocity signal decreases with increasing
orbital period, while large-separation planets must be exquisitely
aligned to transit our line-of-sight from Earth.  In addition, both of
these indirect methods generally require follow-up for one or more
orbital periods, which could be centuries for a long-period exoplanet.

With the development of high-contrast, high angular resolution
imaging, large telescopes can now directly image giant planets at
large distances from their host stars
\cite{marois_et_al2008,kalas_et_al2008,lagrange_et_al2009,kraus+ireland2011}. The addition of spectroscopy to
direct imaging will enable the characterization of giant exoplanet
atmospheres and the development of techniques to find Earth-like
planets that may support life\cite{kawahara_et_al2012}.  This contribution
describes the science case for CHARIS, a new Integral Field
Spectrograph (IFS) designed for taking spectra of exoplanets, to be
built for the Subaru 8.2 m telescope.  CHARIS will
simultaneously obtain spatial and spectral information over the
field-of-view (FOV) by dispersing the entire image on the detector.
CHARIS will be the first high-contrast IFS on an 8m-class telescope in
the Northern Hemisphere, and will achieve an inner-working angle
$\sim$$2 \lambda/D$ and contrasts of up to $10^{-7}$.  CHARIS will
provide both a low-resolution (${\rm R}=14$) mode in which it will
simultaneously collect photons from $\lambda$=0.9-2.4 $\mu$m, and a
high resolution (${\rm R}=65$) mode over a single near-infrared bandpass: $y$-,
$J$-, $H$- or $K$-band.

The technical design of the instrument is described in the accompanying
paper\cite{peters_et_al2012}.
\section{CHARIS Specifications} \label{sec:specifications}

\begin{figure}
\includegraphics[width=\linewidth]{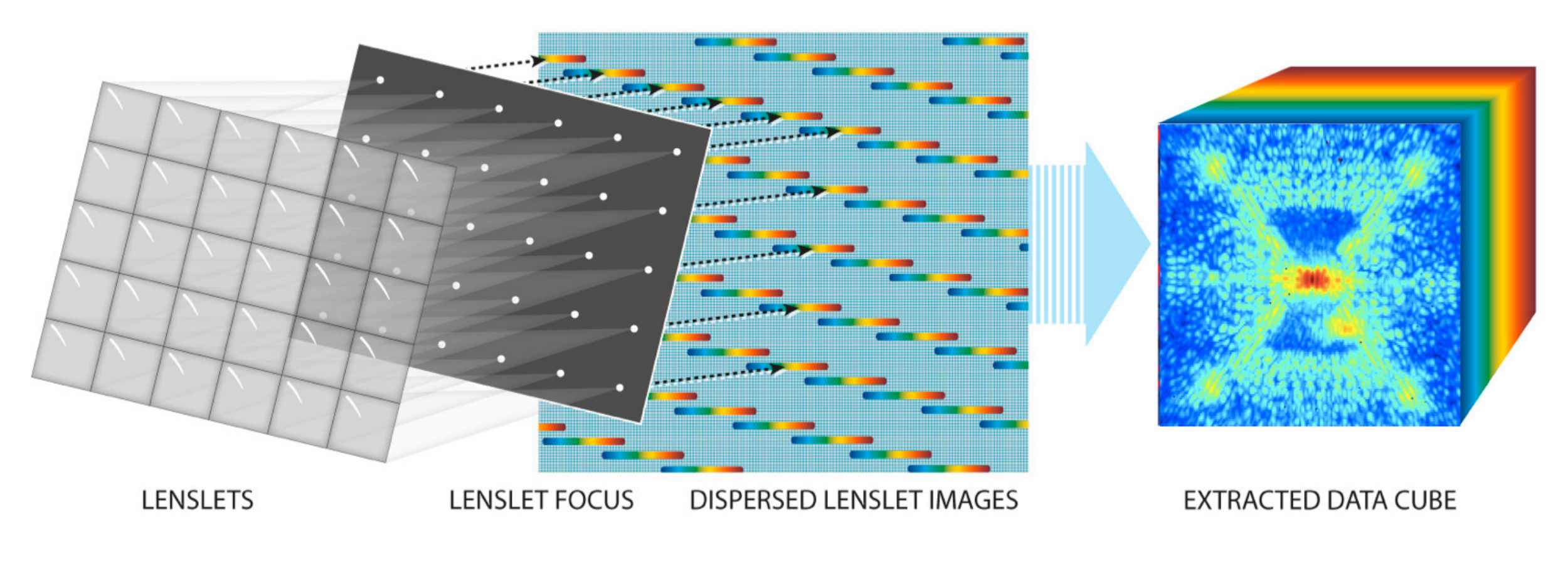}
\caption[example] { \label{fig:datacube} How a lenslet-based IFS
  works. A lenslet array at a focal plane within the instrument
  focuses the incident light into a grid of well-separated points.  
  A pinhole mask at the lenslet's focal plane blocks
  diffracted and scattered light.  Each lenslet's focused light is
  then dispersed onto a Teledyne H2RG detector in such a way that they
  do not overlap.  Finally, a data cube is extracted: a spectrum at each 
  spatial location in the FOV.}
\end{figure} 

CHARIS is a lenslet-based IFS designed for high-contrast imaging and
spectroscopy. The instrument produces a spectrum for each spatial
element (spaxel) and thereby a data cube of the observed field, with
two spatial and one wavelength dimensions (see Figure
\ref{fig:datacube}).  The basic design follows that of the first
astronomical IFS TIGER\cite{bacon_et_al1995} but incorporates
innovations such as specialized lenslets to reduce crosstalk between the 
spectra of neighboring spaxels.  Table \ref{tab:charis_spec} lists the key specifications of
the CHARIS IFS.

CHARIS is part of a third-generation suite of high-contrast
instrumentation at the Subaru telescope, and will be the first
high-contrast IFS on an 8-m class telescope in the Northern
Hemisphere.    Its primary competitor in the Northern Hemisphere will be Project
1640\cite{hinkley_et_al2008} on the Palomar $200''$.  It will complement the similar instruments currently
being built for the Southern 8-m class telescopes: the Gemini Planet
Imager (GPI\cite{macintosh_et_al2008,perrin_et_al2010}) and the Spectro-Polarimetric
High-Contrast Exoplanet Research instrument (SPHERE) on the Very Large
Telescope VLT-Yepun\cite{beuzit_et_al2006,beuzit_et_al2008,claudi_et_al2011}.

\begin{table}
\centering
\caption{CHARIS Specifications \label{tab:charis_spec}}
\smallskip
\begin{tabular}{ l c c c }
Spaxel Scale & \multicolumn{3}{r}{$0.\!\!''0126$ ($2\lambda$/D at 0.9 $\rm \mu m$)} \\
Spectral Range & \multicolumn{3}{r}{0.9 -- 2.4 $\rm \mu m$}\\
Observing Mode & low-R & medium-R & high-R \\
Spectral Resolution& 14 & 33 & 65\\
\smallskip
Bandwidth& 1.5 $\rm \mu m$ & 0.7 $\rm \mu m$ & 0.4 $\rm \mu m$\\
Detector& \multicolumn{3}{r}{$2048 \times 2048$ HgCdTe Hawaii 2RG}\\
Field of View& \multicolumn{3}{r}{$1.\!\!''75 \times 1.\!\!''75$} 
\end{tabular}
\end{table}
\section{High Contrast Instrumentation at Subaru}

\begin{figure}
\centering\includegraphics[width=0.8\linewidth]{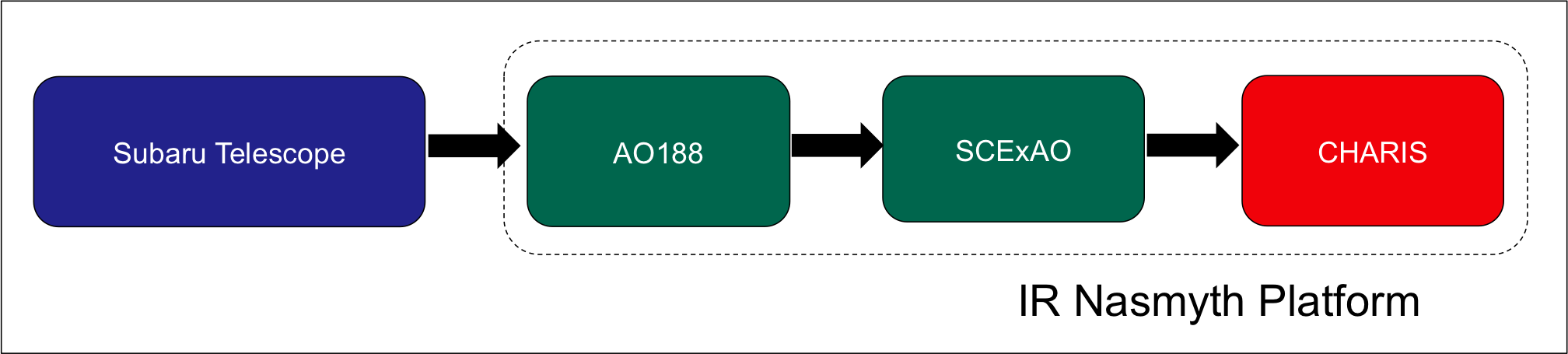}
\caption[example] 
{ \label{fig:nasmyth} 
Layout of the third-generation high-contrast 
instrumentation suite at Subaru's infrared Nasmyth platform.}
\end{figure} 

The Subaru telescope's infrared Nasmyth platform currently
houses a 188-actuator AO system (AO188) and a high-contrast
near-infrared science camera (HiCIAO\cite{suzuki_et_al2010}).  SCExAO, the
Subaru Coronagraphic Extreme AO project\cite{guyon_et_al2011,martinache_et_al2011},
received first light on 5 February 2011 and is currently under
commissioning. When fully finished, SCExAO will provide high-order
wavefront correction with sophisticated coronography using techniques such
as phase-induced amplitude apodization (PIAA\cite{guyon_et_al2005}) and shaped 
pupil coronagraphs (SP\cite{kasdin_et_al2003}).  Further contrast improvements will be
made with focal plane wavefront sensing techniques such as speckle nulling\cite{borde+traub2006}, 
dramatically improving contrast at small inner
working angles. SCExAO will initially feed its beam into HiCIAO for
imaging, immediately improving HiCIAO's performance at small separations
($<$ $0.2''$). The flow chart
in Figure \ref{fig:nasmyth} shows the layout of the third-generation
high-contrast instrumentation suite at Subaru's infrared Nasmyth
platform, with CHARIS replacing HiCIAO as the science camera.

\section{The SEEDS Project}

\begin{figure}
\centering
\includegraphics[width=0.48\linewidth]{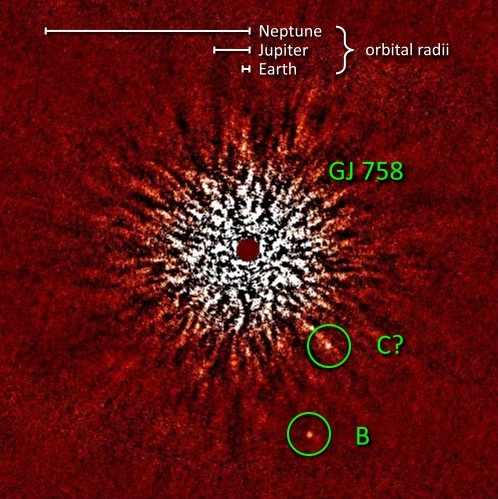}
\includegraphics[width=0.48\linewidth]{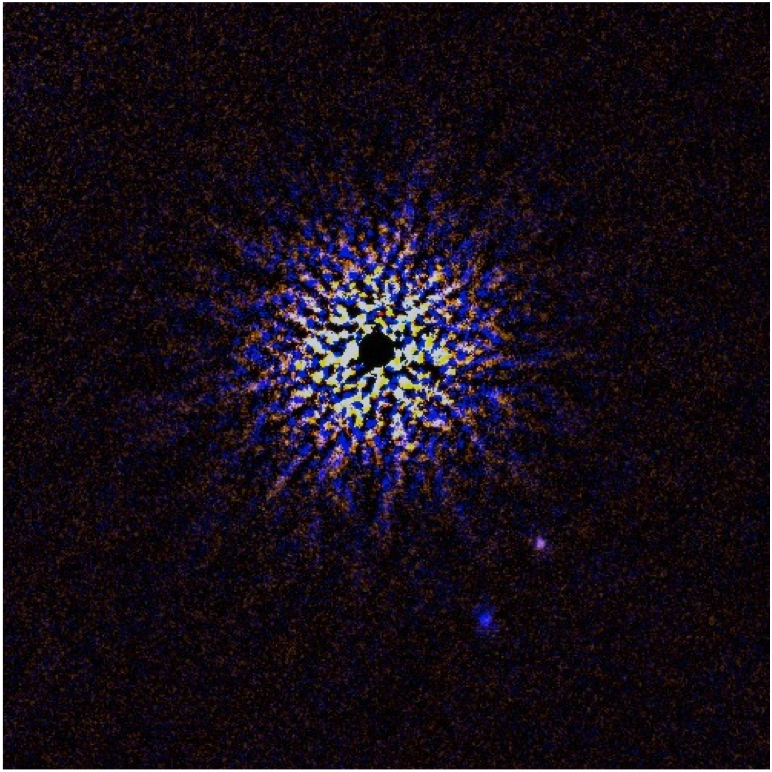}
\label{fig:gj758}
\caption[example] {Discovery of the cold substellar
  object GJ 758 B. {\it Left image}: Subaru AO188+HiCIAO ADI $H$-band
  observation\cite{thalmann_et_al2009}. {\it Right image}: Combined SEEDS AO188+HiCIAO SDI image of
  GJ~758 in CH$_{4}$-on (red) and CH$_{4}$-off (blue), showing the spectral information obtained in adjacent narrow-band
  filters (R.  Kandori et al., in preparation).  GJ 758 B shows
  dramatically different fluxes in and out of the methane band,
  represented here by red and blue, respectively.  The background
  star (indicated by ``C?'' in the left image and confirmed as a
  background source by a common proper motion test) has similar fluxes in CH$_{4}$-on and CH$_{4}$-off,
  showing no evidence of a methane feature.  SDI and IFS data products are capable of detecting and spectrally characterizing detections in a single observation.}
\end{figure}

SEEDS, the Strategic Exploration of Exoplanets and Disks with Subaru,
led by P.I.~Motohide Tamura, is the first of Subaru's strategic
programs.  The project uses HiCIAO for high-contrast imaging in the $H$-band (1.6 $\rm \mu m$).  SEEDS will ultimately survey $\sim$500 nearby
stars to detect and characterize planets and disks around
them\cite{tamura2009}. SEEDS has already discovered substellar
companions of several Jupiter masses around young nearby stars, one
example of which is shown in Figure \ref{fig:gj758}.  HiCIAO operates
in several modes:

\begin{enumerate}
\item Polarimetric differential imaging (PDI) to detect and map the
scattered light from planet-forming disks around young stellar objects
(YSOs)\cite{thalmann_et_al2010,hashimoto_et_al2011,muto_et_al2012} and from debris disks;
\item Angular differential imaging (ADI) to search for point sources,
i.e.,~self-luminous giant planets; and
\item Spectral differential imaging (SDI), in which HiCIAO simultaneously images
	in four narrow sub-bands that straddle the $H$-band CH$_{4}$ feature.
 SDI is also used to provide higher contrast in the search for point sources.
\end{enumerate}
Most of the SEEDS work to date has been carried out in the PDI and ADI
modes; see for example the detection of a cold substellar companion to
the nearby G8 star GJ 758\cite{thalmann_et_al2009,janson_et_al2011}. The SDI mode has
been deployed occasionally for follow-up observations (e.g.,~R. Kandori
et al., in preparation).  HiCIAO SDI splits the bands in intensity with a Wollaston prism, and then passes the beams through either two or four narrow sub-bands that straddle the $H$-band CH$_{4}$ feature.  In this mode, each SDI
narrow band has only $\sim$10\% of the bandpass of the broadband H filter.  HiCIAO's SDI mode thus gains spectral information at the cost of 90\% of the incident $H$-band photons.

Figure \ref{fig:gj758} shows the ADI and SDI images of GJ 758 and its
companion, GJ 758 B.  This is a precursor for the images that CHARIS
will produce.  By dispersing all of the incident light in its
low-spectral-resolution mode, CHARIS will avoid HiCIAO's tradeoff
between spectral information and sensitivity to faint sources.  The
spectral data provided by CHARIS will allow both better suppression of
speckle noise and detailed characterizations of exoplanet atmospheres,
including those of exoplanets discovered by SEEDS.

\section{CHARIS Field-of-View and Spectral Resolution}

\begin{figure}
\centering
\includegraphics[width=0.32\linewidth]{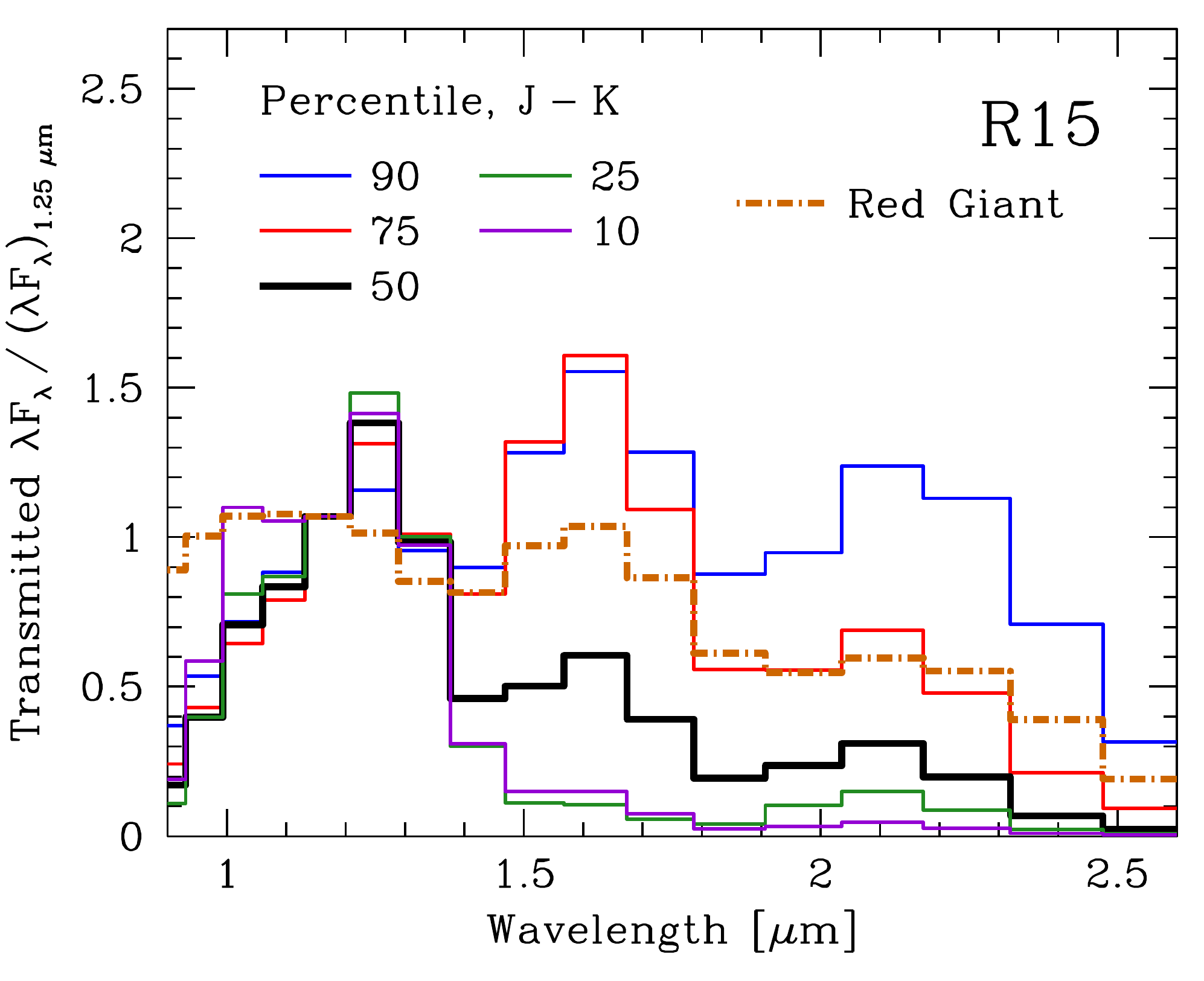}
\includegraphics[width=0.32\linewidth]{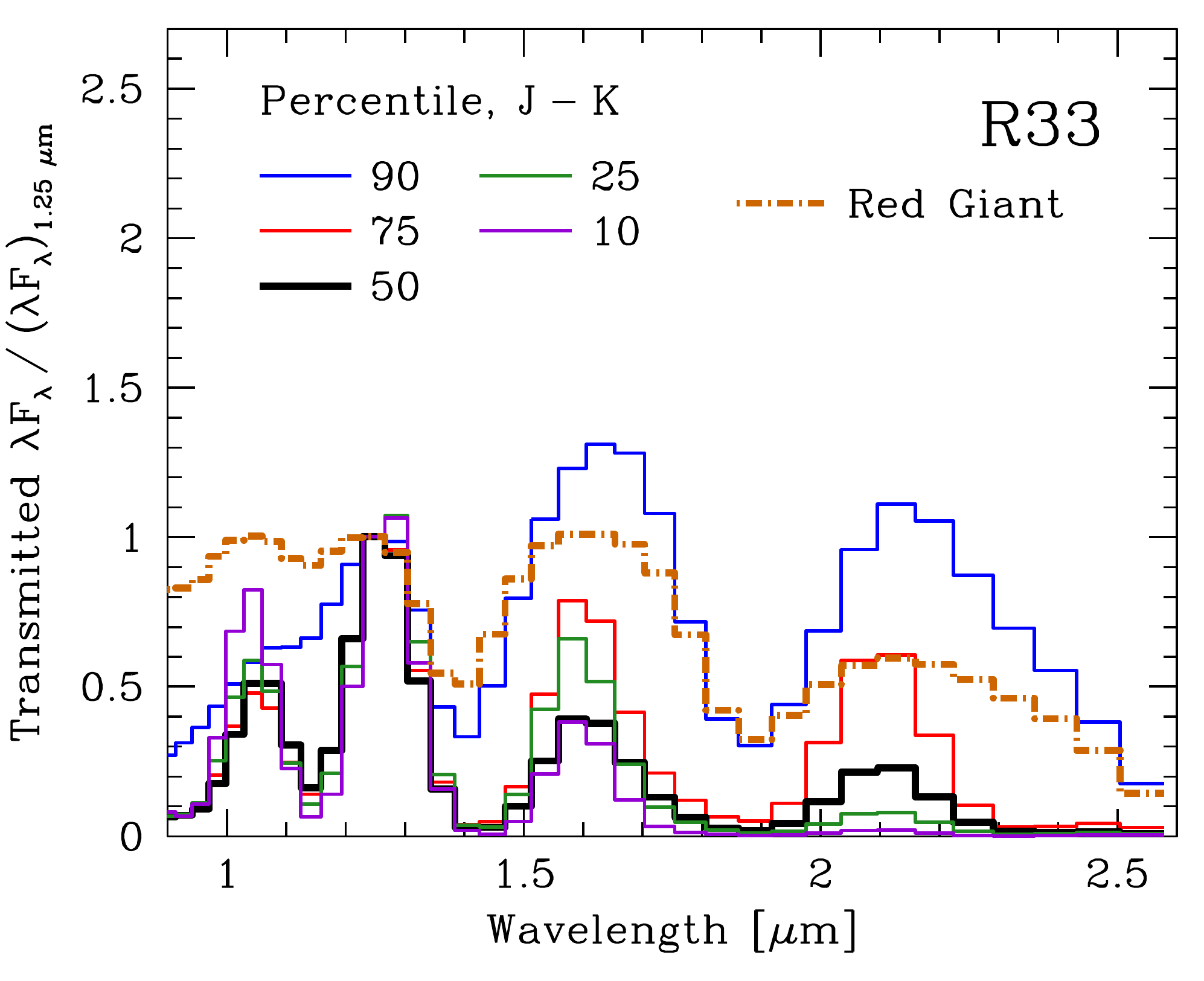}
\includegraphics[width=0.32\linewidth]{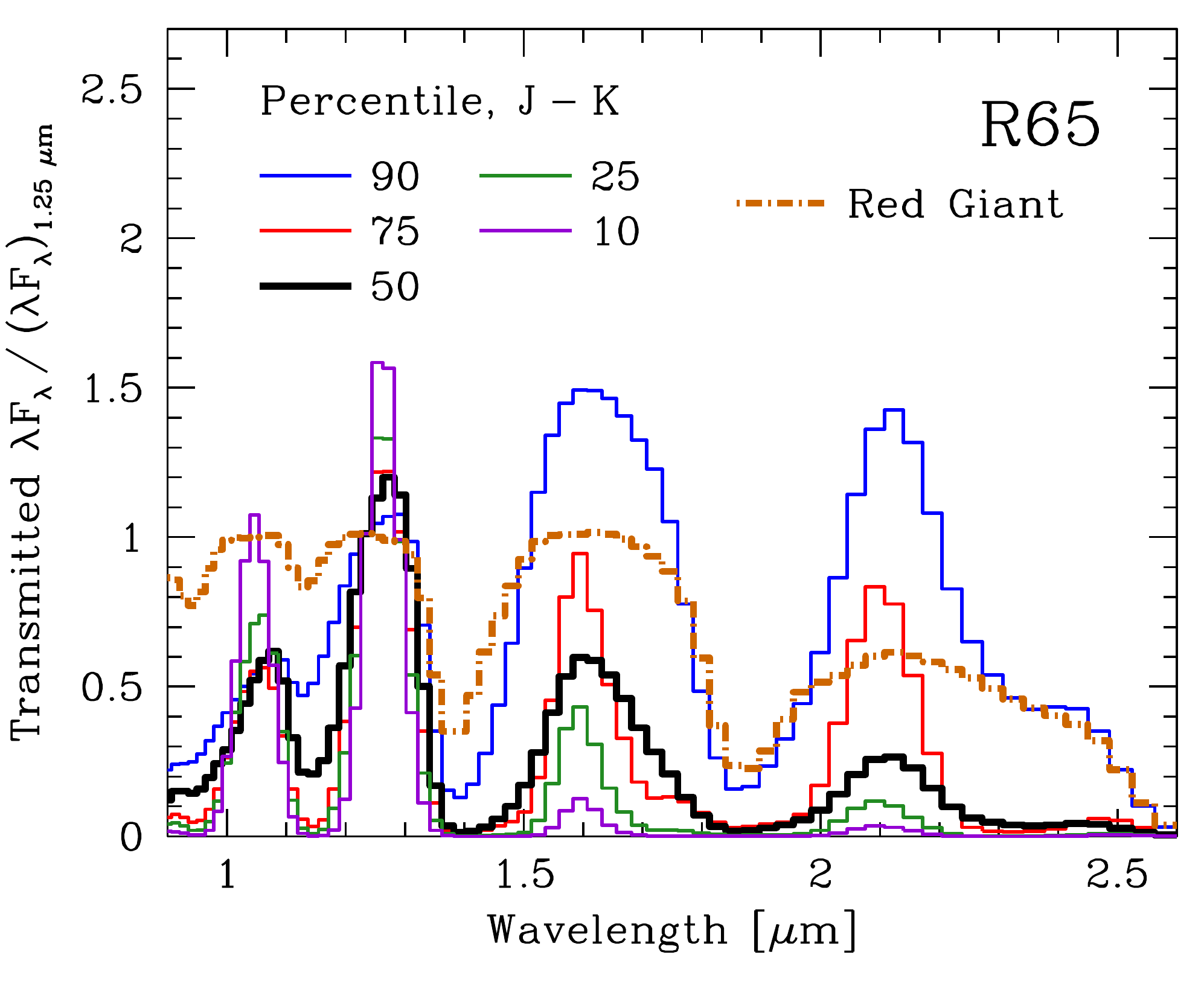}
\caption[example] {\label{fig:resolution} Model exoplanet spectra with
  a range of $J-K$ spectral indices, convolved with the atmospheric
  transmission at Mauna Kea and scaled to the count rate at the CHARIS
  detector.  We overplot a relatively featureless red giant spectrum
  for comparison.}
\end{figure}

\begin{figure}
\centering
\includegraphics[width=0.32\linewidth]{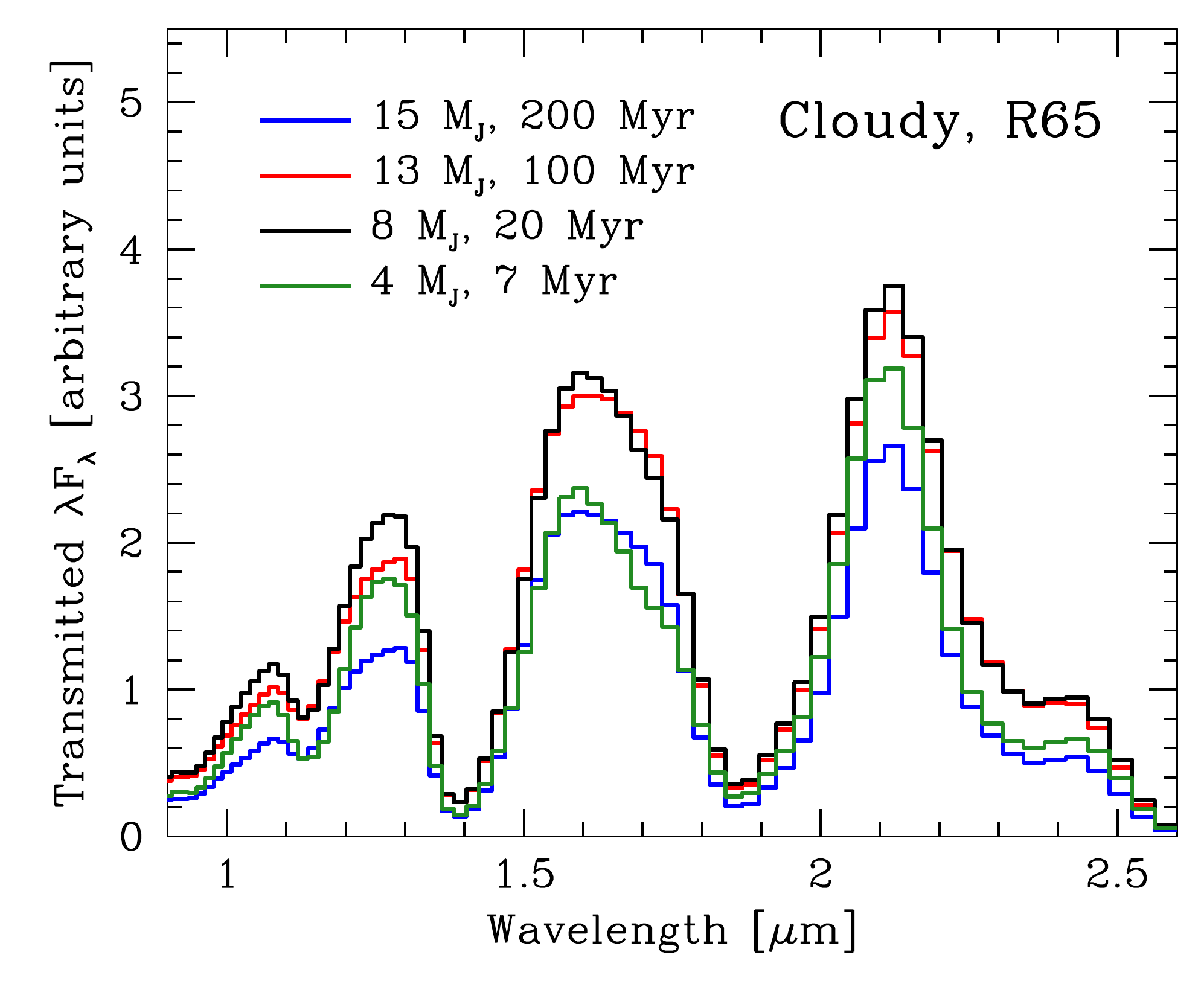}
\includegraphics[width=0.32\linewidth]{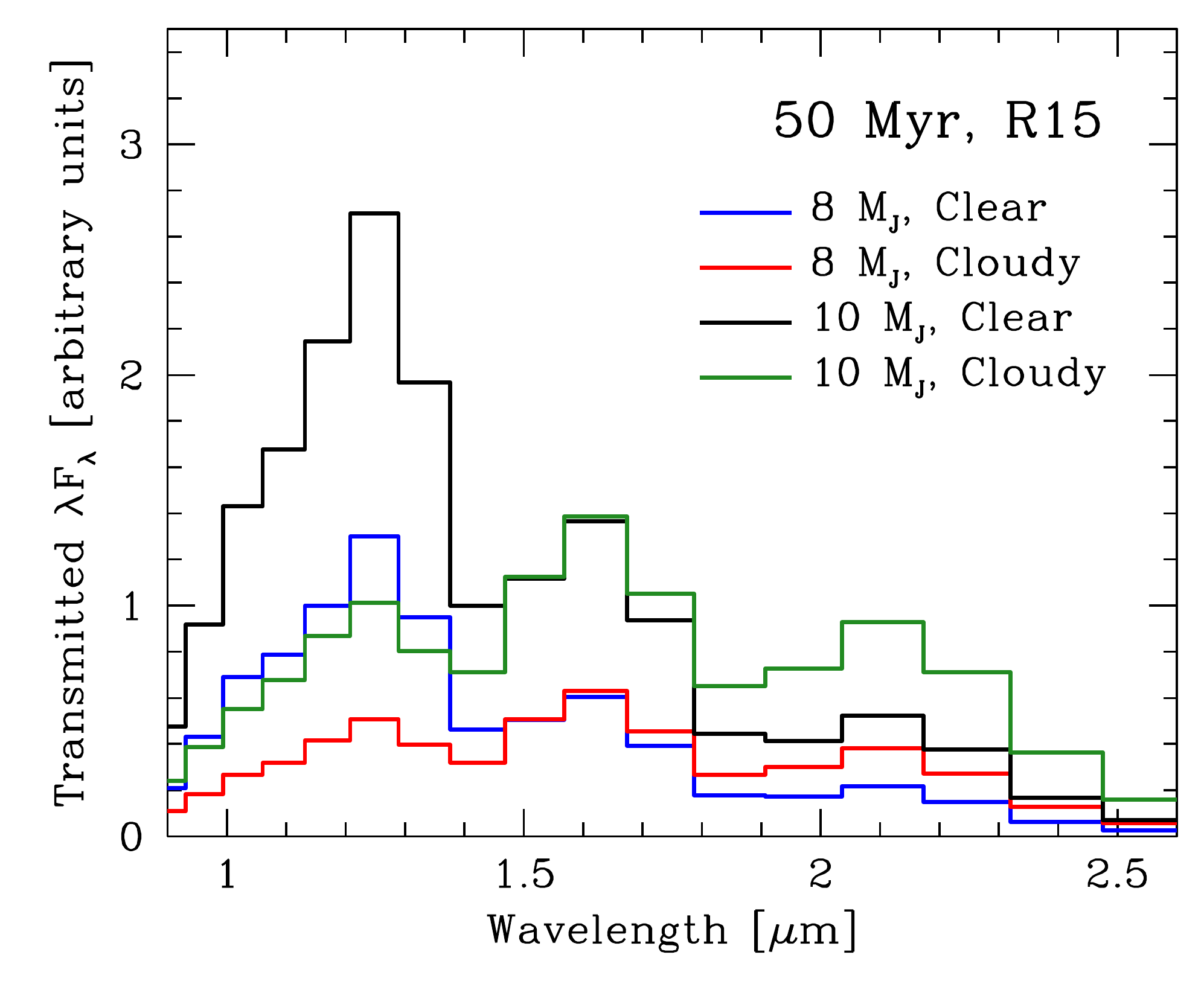}
\caption[example] { \label{fig:spect1} {\it Left panel}: Model exoplanet
  spectra\cite{spiegel+burrows2012} as observed with the CHARIS high spectral
  resolution mode, showing the strong degeneracy between age and
  mass. {\it Right panel}: If a planet has a well-determined age, even
  low-resolution spectra provide strong constraints on the mass and
  insights into atmospheric properties.  The planet and its
  host star are assumed to
  be coeval; observations of the host star provide the age
  constraints. }
\end{figure} 

CHARIS will accept its beam from the extreme AO system, SCExAO, which
will produce a high Strehl ratio (and hence, high contrast) over a
wavelength-dependent radius of about $1''$.  This, the detector size,
and the sampling at the shortest wavelength (the $y$-band at 0.9 $\rm
\mu m$), set our FOV to be $\sim$$1.\!\!''75 \times 1.\!\!''75$ (Table
\ref{tab:charis_spec}).  At 10 pc, $1''$ corresponds to a distance
from the star of 10 A.U., so CHARIS with SCExAO will probe planetary
system scales similar to those of the giant planets in the Solar
system.

The selection of spectral passbands and resolution has been informed
both by recent spectroscopic observations of free-floating substellar
objects (the brown dwarfs), which serve at least in part as proxies
for the properties of giant exoplanets\cite{burgasser2009}, and by detailed models of
giant-planet atmospheres\cite{spiegel+burrows2012}.  Figures
\ref{fig:resolution} and \ref{fig:spect1} illustrate these models for
exoplanets of a range of masses, ages, and assumed initial conditions.
According to these models, the spectra in the $K$- and $H$-bands provide
the best diagnostics of the planetary initial conditions.  Two
mechanisms have been proposed to form giant, large-separation
exoplanets: the core-accretion model\cite{pollack_et_al1996}, and the gravitational
instability model\cite{boss1997}.  These two models predict
different thermodynamic conditions in the newly formed exoplanets;
spectra covering the full wavelength range from 0.9 to 2.4 $\mu$m,
together with well-determined system ages, will thus be able to
discriminate between them.  We therefore decided to include the $K$ band
(2.2 $\rm \mu m$) in CHARIS despite the potential for thermally induced noise.

The current conceptual design for CHARIS covers the near-infrared from
the $y$- to the $K$-band. Our baseline design provides a spectral
resolution of ${\rm R}=33$ with a bandpass of $\Delta\lambda = 0.7$
$\mu$m ($J+H$ band or $H+K$ band). The calculations and simulations shown
in Figures \ref{fig:resolution} and \ref{fig:spect1} suggest two
additional modes: 
\begin{enumerate}
\item ${\rm R} = 15$ with a $\Delta\lambda$ = 1.4 $\mu$m bandpass,
  allowing CHARIS to measure the full wavelength range from the $y$- to
  the $K$-band simultaneously (left panel of Figure \ref{fig:spect1}); and
\item ${\rm R} = 65$ with a $\Delta\lambda$ = 0.4 $\mu$m bandpass,
  which allows higher-resolution observations in each of the bands and
  the measurement of important atmospheric spectral diagnostics such as
  H$_{2}$O, CH$_{4}$, and CO (right panel of Figure \ref{fig:spect1}).
\end{enumerate}
CHARIS will change its spectral resolving power by switching
dispersive elements and changing its bandpass filter. All three
configurations will contain $140 \times 140$ spaxels and 16 spectral
measurements. In all three modes, the plate scale at the lenslet
array will be 12.6 milli-arcseconds (mas, 0$\farcs$001), such that the
shortest wavelength of $\lambda$ = 0.9 $\mu$m is Nyquist
sampled. Details are given in the companion paper\cite{peters_et_al2012}.

\section{CHARIS Data Products}

\begin{figure}
\centering
\includegraphics[width=0.7\linewidth]{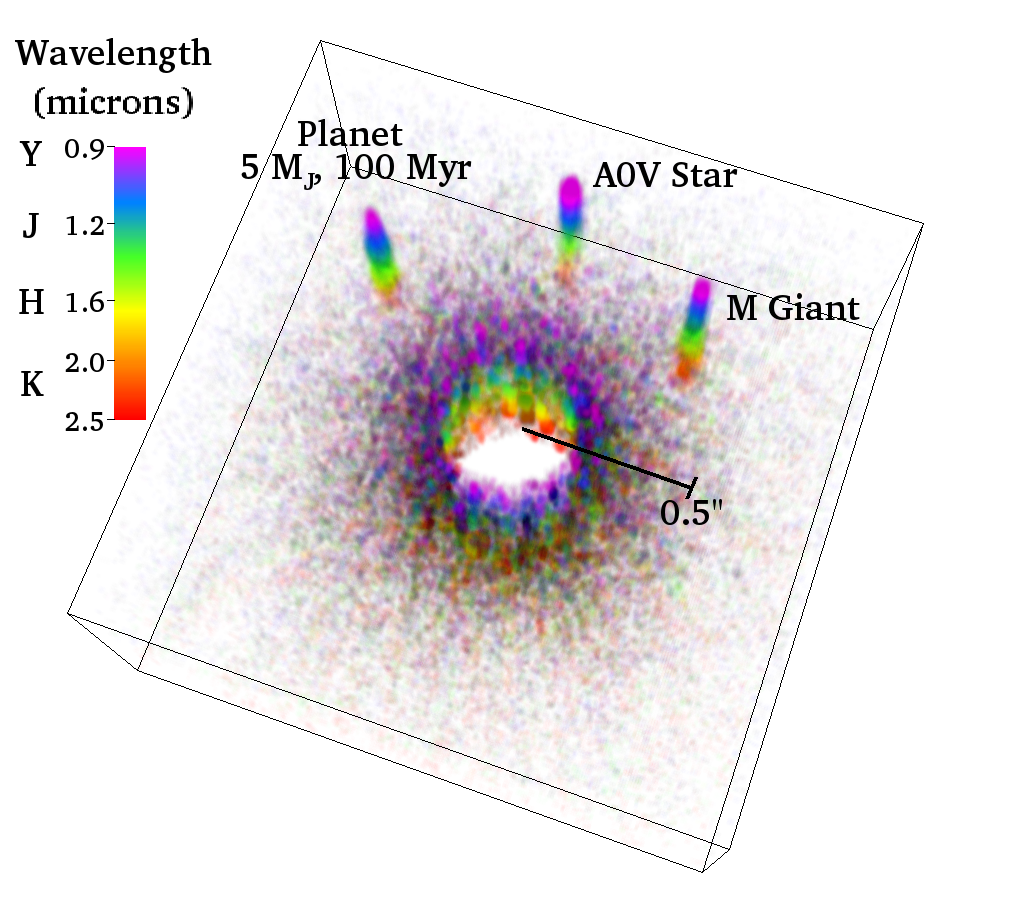}
\caption[example] { \label{fig:simcube} A simulated
  low-spectral-resolution CHARIS datacube showing three point sources
  in the observed field of a solar-type star aged $10^8$ years and at
  a distance of 10 pc.  The point sources are all $10^6$ times fainter
  than the central star and are at a spatial separation of
  $\sim$$0.5''$.  The 5 $\rm M_J$, 100 Myr planet clearly displays spectral
  features, including a much lower flux at short wavelengths, which
  distinguish it from the background stars.  The two background
  objects are an A0V star and an M giant, both of which have
  relatively featureless spectra in the near-infrared.  }
\end{figure} 

A CHARIS observation will produce a data cube of the FOV with one
spectral (\textit{$\lambda$}) and two spatial (\textit{x}, \textit{y})
dimensions.  Calibration spots produced on
the detector by a coarse diffraction grid in SCExAO will provide 3 mas
astrometry with respect to the target star and spectrophotometric
calibration with respect to the target star's spectral energy
distribution.

Figure \ref{fig:simcube} shows a simulated data cube calculated for a
one-hour observation in low-spectral-resolution mode of a young
solar-type star at a distance of 10 pc.  The FOV includes three faint
objects: a companion giant planet, and two background stars. As
Figure \ref{fig:simcube} shows, a single low-resolution observation
can distinguish between exoplanet and stellar spectra, similar but more efficiently
than in the SDI case (see Figure~\ref{fig:gj758}), which is only sensitive to the
CH$_{4}$ feature.  The exoplanet
has much less flux at short wavelengths than the much hotter stars,
and also has stronger spectral features.  Any background star would be
hot enough to make its near-infrared spectrum nearly featureless.  An
observation like that shown in Figure \ref{fig:simcube} would be
followed up to test for common proper motion and to characterize the
exoplanet's atmosphere using CHARIS' high-spectral-resolution
observing mode.


\section{Sensitivity}

\begin{figure}
\centering
\includegraphics[width=0.45\linewidth]{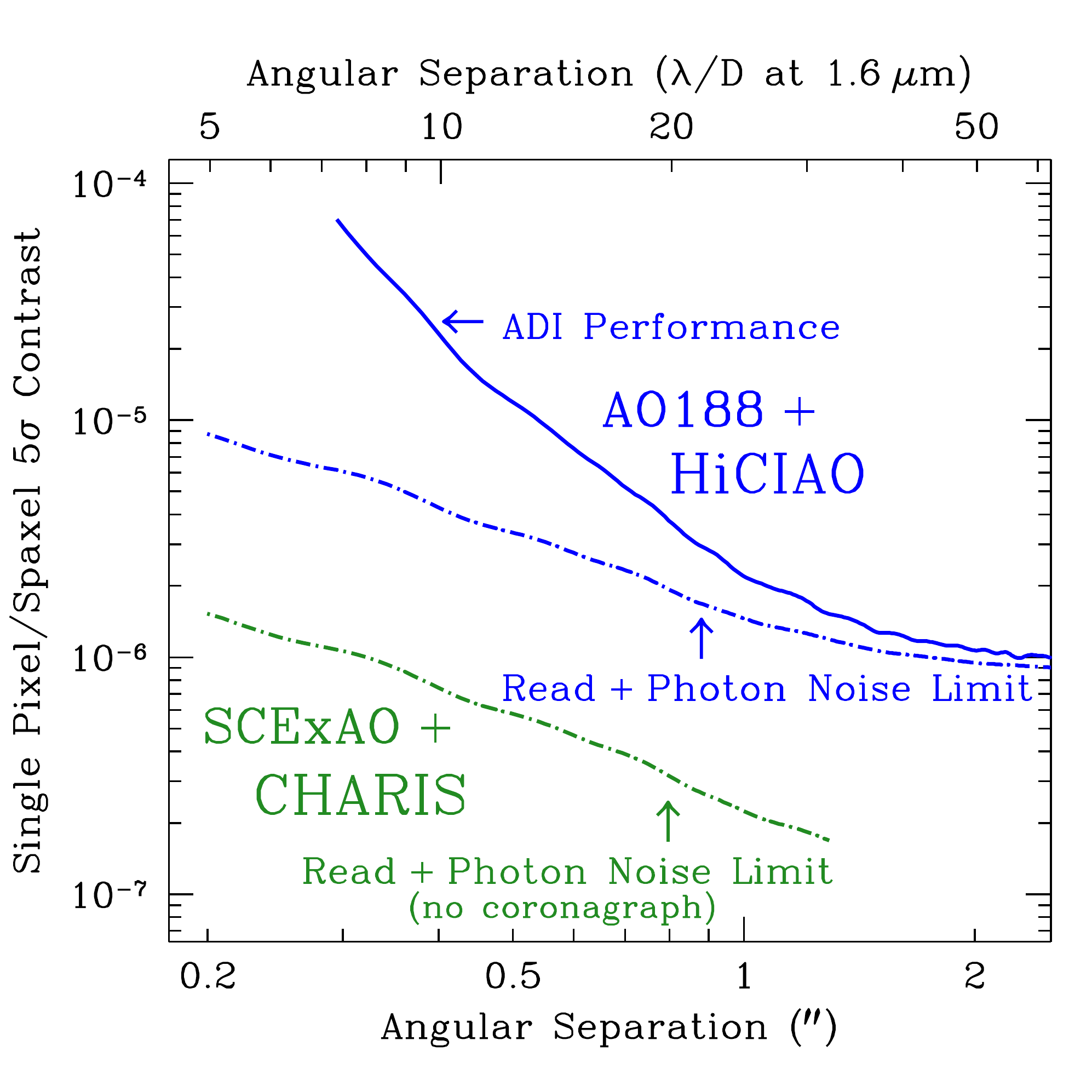}
\caption[example] { \label{fig:contrast} Contrast curve for the
  existing AO188/HiCIAO system (blue) and expected for the combination
  of SCExAO+CHARIS (green).  We show both HiCIAO's limiting performance
  assuming perfect subtraction of the host star and its performance in
  actual observations.  CHARIS' limiting performance will be a factor
  of $\sim$10 better with no coronagraph, and with the advanced
  coronagraphic techniques SCExAO will implement, CHARIS could even
  out-perform the limits shown.  SCExAO + CHARIS will push inward toward 1 $\lambda$/D with the
  PIAA coronagraph.
   }
\end{figure} 

We have estimated the performance of SCExAO+CHARIS relative to that of
AO188+HiCIAO using a one-hour on-sky integration for a fifth magnitude
star as our fiducial observation (Figure \ref{fig:contrast}). SCExAO
will improve the $H$-band Strehl ratio from $\sim$30\% to
$\sim$90\%. However, SCExAO, combined with CHARIS, will suffer a
factor of $\sim$3 loss in throughput relative to AO188+HiCIAO.  On the
other hand, CHARIS will be able to take longer exposures due to the
higher Strehl ratio, increasing the observing efficiency and reducing
the effects of read noise. In its low spectral resolution mode, CHARIS
will collect photons over a much wider bandpass than can HiCIAO.
These factors combine to increase the limiting sensitivity of
SCExAO+CHARIS relative to that of AO188+HiCIAO by a factor of
$\sim$10, assuming perfect subtraction of the host star and no
coronagraph.  

The SCExAO performance shown in Figure \ref{fig:contrast} only
measures the hardware improvements from the AO system and from
CHARIS' spectral coverage.  SCExAO will also implement advanced
coronagraphy and speckle suppression techniques\cite{sparks+ford2002,thatte_et_al2007,mcelwain_et_al2008,
crepp_et_al2011,pueyo_et_al2012}, which
only become possible with the high Strehl ratios it will attain.  In
practice, SCExAO+CHARIS could even out-perform the limits shown.

\section{CHARIS Science}

CHARIS is designed to detect and characterize giant planets in systems
around nearby stars. In combination with SCExAO, it will dramatically
improve contrast relative to AO188+HiCIAO from separations of $\sim
1''$ down to an inner working angle which is essentially Subaru's 
diffraction limit of 2$\lambda$/D ($\sim 0.\!\!''02$,
depending on wavelength).  If the target star has a well-determined
age, even the low-spectral-resolution mode will provide a
well-determined planetary mass.  As Figures \ref{fig:resolution} and
\ref{fig:spect1} illustrate, the atmospheres of giant planets are
expected to show a great deal of spectral diversity due to the effects
of age, metalicity, clouds, and formation conditions. The spectral
characteristics are highly degenerate in age and mass, and the
properties of young planets, which are far easier to detect because
they are brighter, depend in addition on whether the planet was formed
via instabilities in the circumstellar disk (``warm'' start) or via
core accretion (``cold'' start)\cite{spiegel+burrows2012}. 

CHARIS, along with the similar instruments SPHERE and GPI in the
Southern Hemisphere and Project 1640 in the Northern Hemisphere, 
will enable surveys and observations which will
clarify our understanding of the formation and evolution of giant
planets.  Many uncertainties remain regarding exoplanets, from the formation
mechanism itself, to the frequency of exoplanet systems with distant,
giant planets, to the dynamical evolution of such systems.  Data from
IFSs like CHARIS will constrain exoplanet masses and atmospheric
conditions and will uncover the frequency of exoplanet systems, as
well as their dependence on the host star's age and metalicity.  This
will constrain protoplanetary disk models which seek to understand the
formation of planets and sub-stellar objects in the outer disk and the
coevolution of the planets and the disk.

\section{Summary} 
CHARIS will be the first IFS for exoplanet studies on an 8m class
telescope in the Northern Hemisphere.  It will achieve a small inner
working angle ($2 \lambda/D$), and high contrasts of up to $10^{-7}$,
representing a factor of $\gtrsim$10 improvement over the current,
second-generation high-contrast instruments on the Subaru
telescope. CHARIS will provide ${\rm R} = 33$ spectral resolution over
a $1.\!\!''75 \times 1.\!\!''75$ FOV. It will also provide both a
low-spectral-resolution mode (${\rm R}=14$), able to collect imagery
from $0.9-2.4$ $\mu m$, and a high-spectral-resolution mode (${\rm
  R}=65$) over a single near-infrared bandpass.  

CHARIS will offer exceptional sensitivity to detect new exoplanets,
and high contrast and spectral resolution to provide detailed
characterizations of their atmospheres.  It represents a major
advancement in exoplanet science and will help address uncertainties
in the frequency, properties, and formation mechanism of giant
exoplanets.  The instrument should achieve first light at the Subaru 
telescope by the end of 2015.

\acknowledgments       
 
This work was performed with the support of 
the Japanese government's Ministry of Education, Culture,
Sports, Science and Technology through grant-in-aid number
23103002 of the program for Scientific
Research on Innovative Areas.

\bibliography{science_v8.bib}

\end{document}